# Topological properties of reflection zeros of optical differentiators based on layered metal-dielectric-metal structures

Artem I. Kashapov[a,b,*], Evgeni A. Bezus[a,b], Dmitry A. Bykov[a,b], Albert A. Mingazov[a,b], and Leonid L. Doskolovich[a,b]

[a] Image Processing Systems Institute, NRC "Kurchatov Institute", Samara, Russia
[b] Samara National Research University, Samara, Russia
*e-mail: ar.kashapov@gmail.com

**Abstract**— We investigate the topological properties of reflection zeros of three-layer structures consisting of a dielectric layer sandwiched between two metal layers, which can be used as optical differentiators. We show that the reflection zeros possess non-zero topological charges, which makes them topologically protected. With a small perturbation of the parameters of the structure (e.g., a change in one of the layer thicknesses), the reflection zero does not disappear, but shifts in the parameter space, i.e., appears at different wavelength and angle of incidence. We demonstrate that with a further parameter change, two zeros with opposite topological charges (+1 and –1) approach each other, merge, and then disappear. We believe that the obtained results give useful insight regarding the operation of layered metal-dielectric-metal structures possessing reflection zeros.

## 1. INTRODUCTION

In recent years, analog signal processing has again attracted considerable research attention. This is due to the fact that analog computing devices can potentially provide faster operation and lower power consumption than their digital counterparts. Currently, nanophotonics is considered as the most promising platform implementing this approach [1–4]. An important direction in analog optical computing concerns the differentiation of optical signals and the computation of various differential operators in temporal, spatial, and spatiotemporal domains [2, 4–12]. Despite their advantages, analog optical differentiators still do not constitute a reliable alternative to digital processors due to very high sensitivity of the most analog systems to perturbations of structural and signal parameters. In this regard, the development and investigation of optical differentiators being robust to parameter changes is of great interest. One of the promising ways to achieve such an "immunity" consists in using photonic structures exploiting certain topological properties [12–15].

A necessary condition for implementing optical differentiation is the presence of a zero in the reflection or transmission spectrum of the utilized photonic structure [4–12]. One of the simplest classes of optical differentiators operating in reflection is constituted by layered metal-dielectric structures including the "basic" three-layer metal-dielectric-metal (MDM) structure consisting of a dielectric layer sandwiched between two metal layers [16–18]. A reflection zero in the latter structure is associated with a Fabry–Pérot resonance [19, 20] and can be easily obtained by tuning the thicknesses of two upper layers of the MDM structure with all other parameters of the structure and of the incident wave being fixed [16]. In the present work, we theoretically and numerically investigate the topological properties of the reflection zero in such structures, demonstrate its topological protection, and study the behavior of the zeros upon perturbations of the layer thicknesses of the structure.

## 2. GEOMETRY OF THE MDM STRUCTURE AND REFLECTION ZERO CONDITIONS

Geometry of the investigated metal-dielectric-metal structure is shown in Fig. 1(a). It consists of an upper metal layer with thickness $h_{m,1}$ and dielectric permittivity $\varepsilon_{m,1}$, a dielectric layer with thickness $h_d$ and dielectric permittivity $\varepsilon_d$, and a lower metal layer with thickness $h_{m,2}$ and dielectric permittivity $\varepsilon_{m,2}$ placed on a semi-infinite dielectric substrate with dielectric permittivity $\varepsilon_{sub}$. We assume that the medium over the structure has a unit refractive index. In this work, we restrict our consideration to structures with $\varepsilon_{m,1} = \varepsilon_{m,2} = \varepsilon_m$ and $\varepsilon_d = \varepsilon_{sub}$, however, the results and reasoning presented below remain valid also for structures with different materials of the upper and lower metal layers, as well as of the dielectric layer and the substrate.

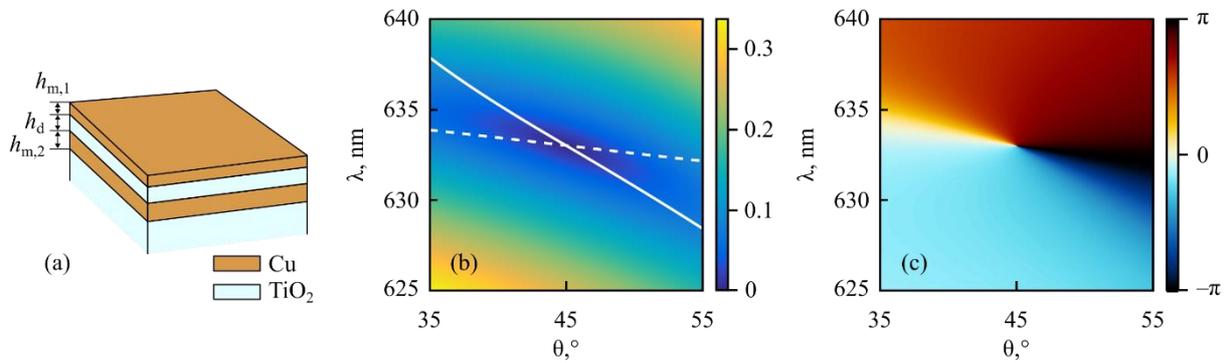

**Fig. 1.** Geometry of the investigated three-layer metal-dielectric-metal structure (a), absolute value (b) and argument (c) of the complex reflection coefficient of an MDM structure with $h_{m,1} = 27.79$ nm, $h_d = 68.08$ nm, $h_{m,2} = 50$ nm possessing a reflection zero at $\lambda = 633$ nm and $\theta = 45°$ vs. wavelength $\lambda$ and angle of incidence $\theta$. Solid and dashed white lines in (b) show the curves, at which the real and imaginary parts of the reflection coefficient vanish, respectively.

For the sake of completeness, following our previous work [16], let us first briefly discuss the conditions of obtaining a reflection zero in the considered MDM structure. The complex reflection coefficient of the MDM structure for a plane wave with a certain linear polarization (transverse electric (TE) or transverse magnetic (TM)) incident on the structure from the superstrate has the form

$$R = \frac{r_1 - \exp(2ik_\perp h_d)\rho(r_1 r_2 - t^2)}{1 - r_2 \rho \exp(2ik_\perp h_d)}, \qquad (1)$$

where $r_1$ and $r_2$ are the complex reflection coefficients of the upper metal layer for the plane waves incident on this layer from above and from below (i.e., from the superstrate and from the dielectric layer), respectively, $t$ is the complex transmission coefficient of this layer (which is the same for the waves incident from above and from below due to reciprocity), $\rho$ is the complex reflection coefficient of the lower metal layer for a plane wave incident from the dielectric layer, $k_\perp = \sqrt{k_0^2 \varepsilon_d - k_\parallel^2}$ is the "vertical" component of the wave vector of the plane waves in the dielectric layer, $k_0 = 2\pi/\lambda$ is the wave number, $\lambda$ is the free-space wavelength, $k_\| = k_0 \sin\theta$ is the in-plane wave vector component of the incident wave, and $\theta$ is the angle of incidence.

As it was mentioned in the Introduction, in order to use a certain photonic structure as an optical differentiator, it is necessary for the reflection (or transmission) coefficient of this structure to vanish. In the considered three-layer MDM structure, one can easily obtain a reflection zero. Indeed, by equating the numerator of Eq. (1) to zero, after simple transformations we arrive at the following conditions of zero reflection ($R = 0$):

$$\left|\frac{r_1}{r_1 r_2 - t^2}\right| = |\rho|, \tag{2}$$

$$h_d = \frac{1}{2k_\perp}\left(2\pi n + \arg\frac{r_1}{\rho(r_1 r_2 - t^2)}\right), \quad n \in \mathbb{N}. \tag{3}$$

In the most general case, the reflection and transmission coefficients of the metal layers $r_1$, $r_2$, $t$, and $\rho$ in Eqs. (2) and (3) depend on multiple parameters, among which are the parameters of the incident wave (free-space wavelength $\lambda$, polarization (TE or TM), and angle of incidence $\theta$) and the parameters of the metal layers (thicknesses $h_{m,1}$ and $h_{m,2}$ and dielectric permittivity $\varepsilon_m$). However, in [16], it was shown that all these parameters except the thickness of the upper metal layer $h_{m,1}$ can be fixed. This means that if we choose a certain wavelength, angle of incidence, and polarization of the incident wave, the materials of the MDM structure, and the thickness of the lower metal layer $h_{m,2}$, then, using Eqs. (2) and (3), we can always find the thicknesses of the upper metal layer $h_{m,1}$ and the dielectric layer $h_d$ providing $R = 0$. The calculation of these thicknesses is performed in the following way. At fixed parameters discussed above, the complex coefficients $r_1$, $r_2$, and $t$ in the left-hand side of Eq. (2) depend only on the thickness $h_{m,1}$, and the right-hand side is constant. Therefore, Eq. (2) can be solved numerically with respect to $h_{m,1}$. Then, the $r_1$, $r_2$, and $t$ values corresponding to the found upper metal layer thickness are plugged into Eq. (3) giving an analytical expression for the thickness of the dielectric layer $h_d$.

## 3. REFLECTION ZERO AND ITS TOPOLOGICAL CHARGE

To illustrate the calculation procedure described above, let us design an MDM structure, for which the reflection coefficient of Eq. (1) vanishes at "central" wavelength $\lambda_0 = 633$ nm and angle of incidence $\theta_0 = 45°$ in the case of TM-polarization of the incident wave. Let us assume the thickness of the lower metal layer of the MDM structure to be $h_{m,2} = 50$ nm. As the dielectric permittivities of the structure, let us use the values $\varepsilon_m = -11.738 + 1.550i$ and $\varepsilon_d = 6.675$, which correspond to copper (Cu) and titanium dioxide (TiO$_2$), respectively [21]. Note that the frequency dispersion of the dielectric permittivities of the materials has been taken into account in all the simulations, the results of which are presented below.

Using Eqs. (2) and (3), a structure with the following thicknesses of the two upper layers was found: $h_{m,1} = 27.79$ nm and $h_d = 68.08$ nm. Figure 1(b) shows the dependence of the absolute value of the complex reflection coefficient of the obtained MDM structure on the free-space wavelength and angle of incidence, which was rigorously calculated using the numerically stable enhanced transmittance matrix approach [22]. Solid and dashed lines in Fig. 1(a) show the zeros of the real and imaginary parts of the reflection coefficient, respectively. It is evident from Fig. 1(b) that the complex reflection coefficient indeed vanishes at the chosen wavelength $\lambda_0 = 633$ nm and angle of incidence $\theta_0 = 45°$, and the zero is isolated in the considered parameter space $(\lambda, \theta)$.

In addition to the absolute value, it is worth studying the argument of the reflection coefficient, which, for the considered example, is shown in Fig. 1(c). This figure shows that around the reflection zero (at which, obviously, the argument is undefined), the argument has a form, which in optics is often referred to as a vortex [23–26]. Speaking more strictly, for an isolated zero of a complex-valued function, we can introduce the so-called topological charge [27, 28]

$$C = \frac{1}{2\pi} \oint d \arg R(\lambda, \theta), \qquad (4)$$

where $d \arg R(\lambda, \theta) = (\partial \arg R / \partial \lambda) d\lambda + (\partial \arg R / \partial \theta) d\theta$ and the integral is taken over a closed contour around the considered zero. From Fig. 1(c), one can clearly see (e.g., by performing numerical integration) that the reflection zero of the designed MDM structure possesses a nonzero topological charge, which equals +1 in the considered case. The existence of a topological charge can also be shown in another way. In the vicinity of the zero, the complex reflection coefficient can be approximated by the following expansion into Taylor series up to the linear terms:

$$R(\lambda, \theta) \approx c_\lambda (\lambda - \lambda_0) + c_\theta (\theta - \theta_0), \qquad (5)$$

where $c_\lambda$ and $c_\theta$ are complex coefficients. For the particular example shown in Fig. 1(b), these coefficients have the following values: $c_\lambda = 33.591 \cdot \exp(2.284i)$ μm$^{-1}$, $c_\theta = 0.576 \cdot \exp(2.920i)$. Let us use the following theorem.

**Theorem.** Let $R(\lambda, \theta)$ be a complex-valued function with a zero at the point $(\lambda_0, \theta_0)$. If

$$\Delta = \begin{vmatrix} \dfrac{\partial \operatorname{Re} R}{\partial \lambda} \bigg|_{(\lambda_0, \theta_0)} & \dfrac{\partial \operatorname{Re} R}{\partial \theta} \bigg|_{(\lambda_0, \theta_0)} \\ \dfrac{\partial \operatorname{Im} R}{\partial \lambda} \bigg|_{(\lambda_0, \theta_0)} & \dfrac{\partial \operatorname{Im} R}{\partial \theta} \bigg|_{(\lambda_0, \theta_0)} \end{vmatrix} \neq 0,$$

then the zero is isolated and it has the associated topological charge with unit absolute value, which equals $C = \operatorname{sgn} \Delta$.

**Proof.** Proven in [29] (see Theorem 6.2 there). □

One can easily show that for a function of Eq. (5), the topological charge of the zero at the point $(\lambda_0, \theta_0)$ is nonzero according to the presented theorem if

$$\arg c_\lambda - \arg c_\theta \neq \pi m, \; m \in \mathbb{Z}. \qquad (6)$$

Obviously, the condition of Eq. (6) is fulfilled for the considered example: $\arg c_\lambda - \arg c_\theta \approx -0.636$. In our previous work, it was shown that it is also generally the case for three-layer MDM structures [18].

An important property of the topological charge of Eq. (4) associated with a reflection zero is that it is an integer value, which can change only in a discrete manner. Therefore, small perturbations of the complex reflection coefficient $R(\lambda,\theta)$ caused, by example, by small changes of one or several thicknesses of the layers cannot change the topological charge. This, in turn, means that the reflection zero will not disappear but move in the parameter space $(\lambda,\theta)$. In this regard, such zeros can be referred to as topologically protected. In the following section, we will numerically investigate the behavior of the reflection zero when one of the parameters of the MDM structure is changed.

## 4. REFLECTION ZERO BEHAVIOR UPON A CHANGE IN A STRUCTURE PARAMETER

Since, as it was shown in the previous section, the reflection zero in the MDM structure is topologically protected, it cannot simply disappear when one of the parameters of the structure is slightly changed. Instead, one could expect one of two behaviors: the reflection zero can either move in the parameter space until one of the parameters becomes non-physical (for example, the angle of incidence exceeds $90°$), or it can "meet" another reflection zero having an opposite topological charge and coalesce with it. Note that the latter behavior has been previously demonstrated for another resonant optical phenomenon, namely, the bound states in the continuum being non-leaky eigenmodes of structures possessing open scattering channels [28, 30, 31].

For the considered structure, let us investigate the behavior of the reflection zero for the case, in which the thickness of the upper metal layer is changed. Fig. 2 shows the absolute values (Figs. 2(a)–2(d)) and arguments (Figs. 2(e)–2(h)) of the reflection coefficient of MDM structures with the thickness $h_{m,1}$ increased by 3.5 nm (Figs. 2(a) and 2(e)), 3.8 nm (Figs. 2(b) and 2(f)), 3.87 nm (Figs. 2(c) and 2(g)), and 3.9 nm (Figs. 2(d) and 2(h)). It is evident from Figs. 2(a) and 2(e) that when the thickness is increased by 3.5 nm, the "initial" reflection zero (see Figs. 1(b) and 1(c)) moves to the point $(\lambda,\theta) = (623.3 \text{ nm}, 65°)$. More interestingly, a second reflection zero located at $(616.8 \text{ nm}, 73.5°)$ "arrives" nearby. From Fig. 2(e), one can see that these reflection zeros possess opposite topological charges (+1 and –1, respectively). When the thickness of the upper metal layer is further increased (see Figs. 2(b) and 2(f)), the zeros move closer to each other and can be found at $(621.3 \text{ nm}, 67.9°)$ and $(616.6 \text{ nm}, 71.5°)$. Then, at a thickness increase of 3.87 nm with respect to the initial value (Figs. 2(c) and 2(g)), the zeros coalesce and form a single reflection zero at the point $(620 \text{ nm}, 69.74°)$. Note that in this case, the curves, at which the real and imaginary parts of the reflection coefficient vanish, do not intersect, but touch each other (see Fig. 2(c)). By calculating the integral of Eq. (4), we find that the resulting reflection zero has zero topological charge, i.e., the topological protection is lifted. In this case, one could expect the disappearance of this reflection zero with a further increase in $h_{m,1}$. This is exactly what happens in Figs. 2(d) and 2(h), which correspond to the thickness increase of 3.9 nm. It is evident that the reflection zero does not exist anymore; in particular, the curves showing the zeros of the real imaginary parts of the reflection coefficient in Fig. 2(d) do not cross. Let us note that, according to the results of additional numerical simulations not presented here for the sake of brevity, a similar behavior is observed when the thickness of one of the other layers of the structure (the dielectric layer or the lower metal layer) is changed.

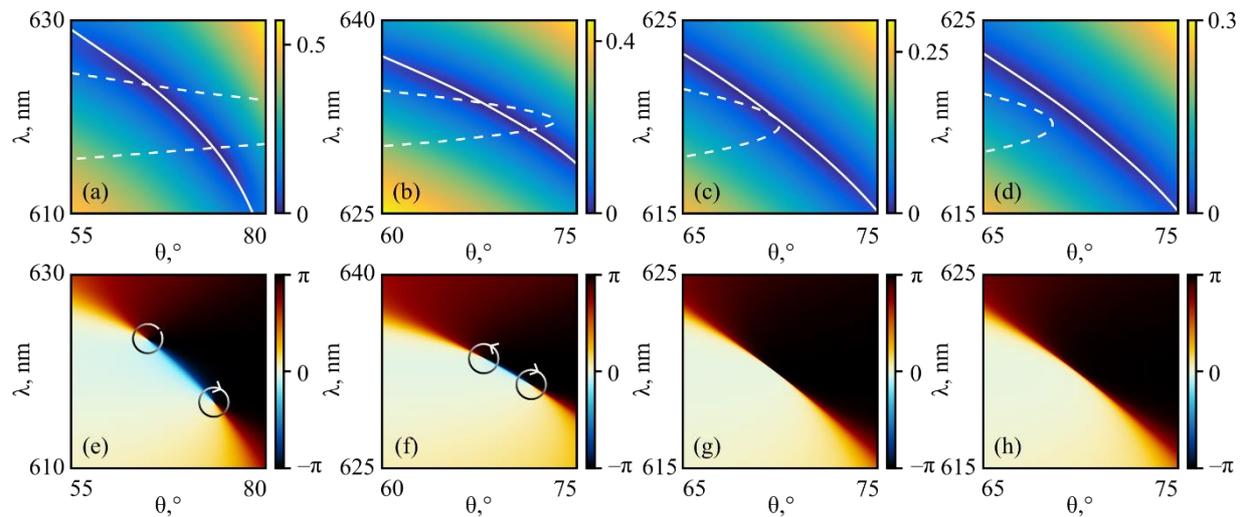

**Fig. 2.** Absolute value (a)–(d) and argument (e)–(h) of the reflection coefficient of MDM structures with the thickness of the upper metal layer increased by 3.5 nm (a), (e), 3.8 nm (b), (f), 3.87 nm (c), (g), and 3.9 nm (d), (h) vs. wavelength $\lambda$ and angle of incidence $\theta$. Solid and dashed white lines in (a)–(d) show the curves, at which the real and imaginary parts of the reflection coefficient vanish, respectively. Circles with arrows around the reflection zeros in (e) and (f) schematically show their topological charges.

## 5. CONCLUSIONS

In the present work, we studied the topological properties of reflection zeros of three-layer structures consisting of a dielectric layer sandwiched between two metal layers. We demonstrated that with these zeros, non-zero topological charges can be associated, which makes them topologically protected. This means that upon small changes in the parameters of the MDM structure, the zeros do not disappear but move in the "wavelength–angle of incidence" parameter space. With a further parameter change, the zeros having opposite topological charges can coalesce and then disappear. We believe that the obtained results will be useful for the design of optical differentiators and filters based on multilayer metal-dielectric structures.

## FUNDING

This work was supported by the Russian Science Foundation (project 24-12-00028).